\documentclass{svproc}
\usepackage{hyperref}
\usepackage{graphicx} 
\usepackage{subfig}
\usepackage{booktabs}
\usepackage{amsmath}
\usepackage{tikz}
\usepackage{cite}
\usepackage{threeparttable}
\usepackage{comment}
\usepackage{xcolor}
\usepackage{ifthen}
\usepackage{float}
\usepackage{changepage}
\usepackage{listings}

\newcommand{\marcom}[2]{%
  \ifthenelse{\equal{#1}{EG}}{ % If #1 is EG
    {\noindent\footnotesize\tt\textcolor{red}{\fbox{#1} #2}} % EG -> Red
  }{ % Else
    \ifthenelse{\equal{#1}{AF}}{ % If #1 is AF
      {\noindent\footnotesize\tt\textcolor{green}{\fbox{#1} #2}} % AF -> Green
    }{ % Else
      \ifthenelse{\equal{#1}{RC}}{ % If #1 is RC
        {\noindent\footnotesize\tt\textcolor{brown}{\fbox{#1} #2}} % RC -> Brown
      }{ % Else (If none match)
        {\noindent\footnotesize\tt\textcolor{blue}{\fbox{#1} #2}} % Default -> Blue
      }
    }
  }
}

\newcommand{\keyw}[1]{“#1”}
\usetikzlibrary{positioning, shapes.geometric, arrows.meta, fit, calc, decorations.pathreplacing}

\usetikzlibrary{positioning, shapes.geometric, arrows.meta, fit, calc}

\tikzset{
  process/.style={rectangle, draw, minimum width=3cm, minimum height=1cm, align=center},
  decision/.style={diamond, draw, minimum width=3cm, minimum height=1cm, align=center},
  database/.style={cylinder, draw, minimum height=1cm, aspect=0.5, align=center},
  io/.style={trapezium, trapezium left angle=60, trapezium right angle=120, draw, align=center},
  line/.style={draw, -{Latex[]}},
  note/.style={draw=none, align=center, font=\itshape}
}

\begin{document}

\mainmatter              % start of a contribution

\title{UK Finfluencers: Exploring Content, Reach, and Responsibility}
\titlerunning{Finance and UK Finfluencers}  % abbreviated title (for running head)

\author{Essam Ghadafi\inst{1} \and Panagiotis Andriotis\inst{2}}
\authorrunning{Essam Ghadafi and Panagiotis Andriotis} % abbreviated author list (for running head)

\tocauthor{Essam Ghadafi and Panagiotis Andriotis}

\institute{
School of Computing, Newcastle University, Newcastle upon Tyne, UK\\
\email{essam.ghadafi@newcastle.ac.uk}
\and
Department of Computer Science, University of Birmingham, Birmingham, UK \\
\email{p.andriotis@bham.ac.uk}
}
            % typeset the title of the contribution

%\title{Finance, Influence, and Responsibility: Examining UK Finfluencers' Content and Network}
%\author{Essam Ghadafi \thanks{School of Computing, Newcaslte University, UK.  Email: %\texttt{essam.ghadafi@newcastle.ac.uk}}  \and Panagiotis Andriotis \thanks{Birmingham University} }
%\date{}

\maketitle

\begin{abstract}
The rise of social media financial influencers (finfluencers) has significantly transformed the personal finance landscape, making financial advice and insights more accessible to a broader and younger audience. By leveraging digital platforms, these influencers have contributed to the democratization of financial literacy. However, the line between education and promotion is often blurred, as many finfluencers lack formal financial qualifications, raising concerns about the accuracy and reliability of the information they share. This study investigates the patterns and behaviours of finfluencers in the UK on TikTok, focusing not on individual actions but on broader trends and the interactions between influencers and their followers. The aim is to identify common engagement patterns and propose guidelines that can help protect the public from potential financial harm. Specifically, the paper contributes a detailed analysis of finfluencer content categorization, sentiment trends, and the prevalence and role of disclaimers, offering empirical insights that inform recommendations for safer and more transparent financial communication on social media.
\end{abstract}
%------------------------------------------------------------------
\section{Introduction}
\label{sec:intro}
The rise of social media has transformed the dissemination of financial information, with so-called financial influencers (or for short finfluencers) playing an increasingly significant role in shaping public investment behaviour. Platforms such as TikTok and Instagram have become key sources of financial content, where influencers promote investment strategies, trading platforms, and cryptocurrency ventures to wide audiences, often without proper regulatory oversight \cite{2023insta,2024tiktokinsta}. While these influencers have the potential to democratise access to financial knowledge, concerns over misinformation, misleading promotions, and the risk of financial harm have drawn growing scrutiny from regulators.

In response to these challenges, the UK's Financial Conduct Authority (FCA) has strengthened its oversight of financial promotions on social media, emphasising that all investment-related content must be fair, clear, and not misleading. Recent enforcement actions, including interviews under caution and legal proceedings against finfluencers, signal a move towards stricter regulation \cite{FCA2024}. This paper analyzes finfluencer content, disclaimers, topics, and audience sentiment to assess how financial advice, marketing, and regulation interact in digital spaces.

TikTok remains the most popular social media platform globally, evidenced by its leading position in mobile application downloads. In 2024, it recorded 773 million downloads, surpassing all other social apps and reinforcing its dominance in the digital ecosystem \cite{businessofapps2024}.
Previous research has examined influencer marketing effectiveness on TikTok, often focusing on entertainment-oriented traits such as humour and originality (e.g., \cite{BARTA2023}), as well as the platform’s structural impact on creative labour and content commodification \cite{Hayes2024}. This study shifts the focus to finfluencers in the UK, analyzing their content strategies, follower interactions, and implications for financial literacy and consumer protection. It identifies behavioural patterns, risks, and communication strategies, offering empirically grounded recommendations for safer, more transparent financial communication. Given TikTok’s popularity among youth \cite{2020Tiktokpopular}, examining finfluencers—whose financial advice may lead to harm or regulatory breaches—is vital for safeguarding financial well-being and ensuring proper oversight of social media content.

\subsection{Our contribution}
We contribute a novel TikTok dataset and associated analyses, covering 71 UK-based finfluencer accounts active between April and September 2024. The dataset includes metadata and extracted text (e.g., descriptions, transcripts) from 13,216 videos, along with 104,097 viewer comments. Our analyses examine engagement and topic trends, the follower graph among finfluencers, the presence of disclaimers in videos and bios, and sentiment in viewer comments.

\subsection{Paper Organization}
The paper is organized as follows: Section~\ref{sec:relatedwork} discusses related work, Section~\ref{sec:methodology} covers the methodology, Section~\ref{sec:findata} analyzes finfluencer data, Section~\ref{sec:Videos} discusses video and engagement analysis, and Section~\ref{sec:summary} concludes the paper.
~\\

{\noindent\textbf{Acknowledgment.}} 
The authors gratefully acknowledge funding from the {National Research Centre on Privacy, Harm Reduction and Adversarial Influence Online (REPHRAIN)}.

%\include{Introduction.tex}
%------------------------------------------------------------------
%##########################################################################################
%##########################################################################################
%##########################################################################################
\section{Related Work}
\label{sec:relatedwork}
Promotions and giveaways are commonly used by influencers to grow their follower base, but these same tactics are also exploited by cybercriminals to attract victims and drive the sale of premium illicit content~\cite{2025darkgram}. In the context of cryptocurrency fraud, regulatory bodies have identified investment scams as one of its most widespread forms~\cite{web25}. These scams often involve professional-looking websites that promise unrealistically high returns to lure unsuspecting investors~\cite{web25}. Prior research also suggests that influential online figures can be leveraged to boost attention toward specific investment opportunities~\cite{nft2024}. Beyond giveaway scams, the literature has documented a wide array of cryptocurrency-related frauds, including high-yield investment scams, pump-and-dump schemes, fake or risky crypto services, initial coin offering (ICO) scams, fraud in non-fungible token (NFT) markets, and phishing attacks themed around cryptocurrencies~\cite{imc24}. These findings underscore the importance of identifying the motivations and credibility of influential actors who disseminate financial advice on social platforms.

Recent studies emphasize the expanding influence of social networks on personal financial activities, particularly among the Millennial and Gen Z populations~\cite{cao2020social}. Baird~\cite{Baird2023} explores how these digitally native generations use platforms for peer-to-peer (P2P) payments, crowdfunding, social commerce, and financial education. With 85\% of people aged 18 to 44 years using P2P payment apps such as Venmo, which incorporate social networking features, social networks are becoming essential tools for facilitating financial transactions. Moreover, crowdfunding initiatives are frequently coordinated through social networks, employing hashtags and online communities to boost participation. However, these developments also pose risks, including financial misinformation and fraud. According to Sortlist Data Hub~\cite{sortlist2022}, 36\% of young adults get financial advice from social media.

Teens increasingly engage in cryptocurrency despite age restrictions, often using their parents' accounts~\cite{Bouma-Simsetal2024}. Social media contributes to this trend by exposing minors to influencers who promote crypto as a profitable venture. An analysis of 1,676 Reddit posts in teen communities shows they are motivated by short-term gains, entertainment, ideology, or technological interest—motifs often echoed by finfluencers~\cite{Bouma-Simsetal2024}.

Researchers~\cite{2024beware} propose using finfluencer quality indicators to establish a ``Finfluencer Quality Score'' aimed at identifying outliers and bad actors in the financial influencer space. This score would help distinguish trustworthy influencers from those promoting misleading advice or manipulative behavior. Other studies, e.g.\cite{2024retail}, examine the impact of finfluencers on retail investment. Their findings show that finfluencers, particularly those with large followings, significantly affect portfolio decisions and trading behavior among their followers.

Multidisciplinary studies such as~\cite{singh2024rise} attempt to investigate the extent to which finfluencers can be considered experts capable of acting as information intermediaries contributing to the efficiency of financial markets. In general, when researchers considered short time frames for studying finfluencers' financial recommendations and analysis, they found that these individuals should not be considered experts, particularly because they often endorse risky investments.

Scholars in Nepal~\cite{decision2025} study factors that affect financial decision making and focus on investors' attitudes. They advocate that expertise, content quality, and credibility have significant predictive power to explain attitudes toward investment. Others take a global approach~\cite{impact25}, focusing on mega influencers. They find that mega influencers impact investor attention, volatility, and trading volume, but not stock returns. Only top influencers with extreme sentiment posts can affect returns, though the effect is short-lived.

Gerritsen and Regt~\cite{consu2025} find that stocks and cryptocurrencies recommended by finfluencers often see negative returns post-recommendation. They suggest that influencers' endorsements are driven by social heuristics, posing risks for investors who follow them. Therefore, it is suggested that stricter regulatory oversight is needed to ensure transparency and protect consumers' financial well-being. Krause~\cite{2025crypto} examines the impact of finfluencers on crypto markets, highlighting systemic risks and regulatory challenges. He notes harmful activities, such as pump-and-dump schemes and undisclosed promotions, and reveals negative returns from following influencers' recommendations, underscoring the inadequacy of current regulations.

Bongini et al.~\cite{2025investors} highlight social media's significant role in shaping market movements, particularly in stocks. They find that posts from influential users can greatly impact stock returns and trading volumes, emphasizing social media's influence on financial market behavior.

Eynde and Rogge~\cite{2023legal} discuss the legal implications of disclaimers used by finfluencers, stressing the need for caution when offering financial advice. Disclaimers are essential for transparency, avoiding miscommunication, and managing expectations. Additionally, authors in~\cite{real2025} explore strategies finfluencers use to emphasize authenticity, such as transparency, immediacy, ordinariness, and passion, fostering stronger follower relationships. They also highlight the limitations of international research on social media influencer marketing.

Other researchers~\cite{young2023} also stress the rise of finfluencers and their popularity among young audiences and they conduct content analysis on social media posts aiming to realise how people consume financial recommendation content. They find that influencers with financial accreditation generate a more positive affective response when they are compared with influencers who share information stemming from their personal experience. They additionally comment of consumers' affective responses related to influencers’ gender and race. Others~\cite{fin2024} conclude that explicit financial literacy with balanced recommendations that shape consumer financial behavior could potentially amplify the role of financial influencers in financial decision making. In addition, they highlight the importance of transparency, authenticity, and accuracy in influencer content to enable responsible spending decisions.

Finally, Hayes and Ben-Shmuel~\cite{Hayes2024} uncover how finfluencers transform individuals' engagement with personal finance through social media. They emphasize the need for ongoing research to promote financial literacy and ensure that the benefits of this form of financial influence are distributed equitably, while also mitigating potential harms.

%------------------------------------------------------------------
%------------------------------------------------------------------
\section{Methodology}
\label{sec:methodology}
This section outlines the methodology employed in this study. 
\subsection{Ethical Considerations}
\label{sec:ethicalapproval}
Prior to data collection, ethical approval was obtained from the authors' universities and the project’s funding body. Data was collected using TikTok’s official Research API~\cite{TikTokResearchAPI}, with access and approval granted by TikTok. All research stages adhered to ethical guidelines, TikTok’s terms of service, and respected privacy, responsible data use, and academic integrity.
\subsection{Data Collection}
\label{sec:Datacollected}
To classify finfluencers, we gathered 10,000 videos for each month from April to September 2024
 using a \texttt{curl} command executed from Python to query the TikTok endpoint:  \url{https://open.tiktokapis.com/v2/research/video/query/}.
 
We searched for videos containing any of the following keywords in their \texttt{hashtag\_name} and \texttt{keyword} fields:
%\begin{lstlisting}
\textquotedblleft budgeting\textquotedblright, \textquotedblleft crypto\textquotedblright, \textquotedblleft cryptocurrency\textquotedblright, \textquotedblleft debtfree\textquotedblright, \textquotedblleft debtpayoff\textquotedblright, \textquotedblleft finance\textquotedblright, 
\textquotedblleft financetiktok\textquotedblright, \textquotedblleft financialeducation\textquotedblright, \textquotedblleft financialfreedom\textquotedblright, \textquotedblleft financialliteracy\textquotedblright, 
\textquotedblleft fintok\textquotedblright, \textquotedblleft investing\textquotedblright, \textquotedblleft investingtips\textquotedblright, \textquotedblleft money\textquotedblright, \textquotedblleft moneygamemoneymindset\textquotedblright, 
\textquotedblleft moneymanagement\textquotedblright, \textquotedblleft moneytok\textquotedblright, \textquotedblleft passiveincome\textquotedblright, \textquotedblleft personalfinance\textquotedblright, 
\textquotedblleft realestateinvesting\textquotedblright, \textquotedblleft retirementplanning\textquotedblright, \textquotedblleft richmindset\textquotedblright, \textquotedblleft savemoney\textquotedblright, 
\textquotedblleft stockmarket\textquotedblright, \textquotedblleft stocks\textquotedblright, \textquotedblleft wealth\textquotedblright, \textquotedblleft wealthbuilding\textquotedblright.
%\end{lstlisting}
Additionally, we filtered for videos whose \texttt{region\_code} field was set to {\textquotedblleft GB\textquotedblright }. 

We identified users who had a total of at least 50 comments on their videos and who posted at least one video in each of the specified months. This yielded 71 users for analysis. We then collected 13,215 videos belonging to those 71 users during those months, including 4,650 with spoken transcripts in the \texttt{voice\_to\_text} field, and 104,097 comments on their videos.
\begin{comment}
Comments per month
# September 20851
# August 23642
# July 13603
# June 9684
# May 11897
# April  15345
\end{comment}

\section{Finfluencers Data Analysis}
\label{sec:findata}
This section provides a comprehensive examination of finfluencer engagement and activity.
\subsection{Finfluencers Activity and Engagement} 
\begin{figure}[H]
    \centering
    \subfloat[Likes]{%
        \includegraphics[width=0.32\textwidth]{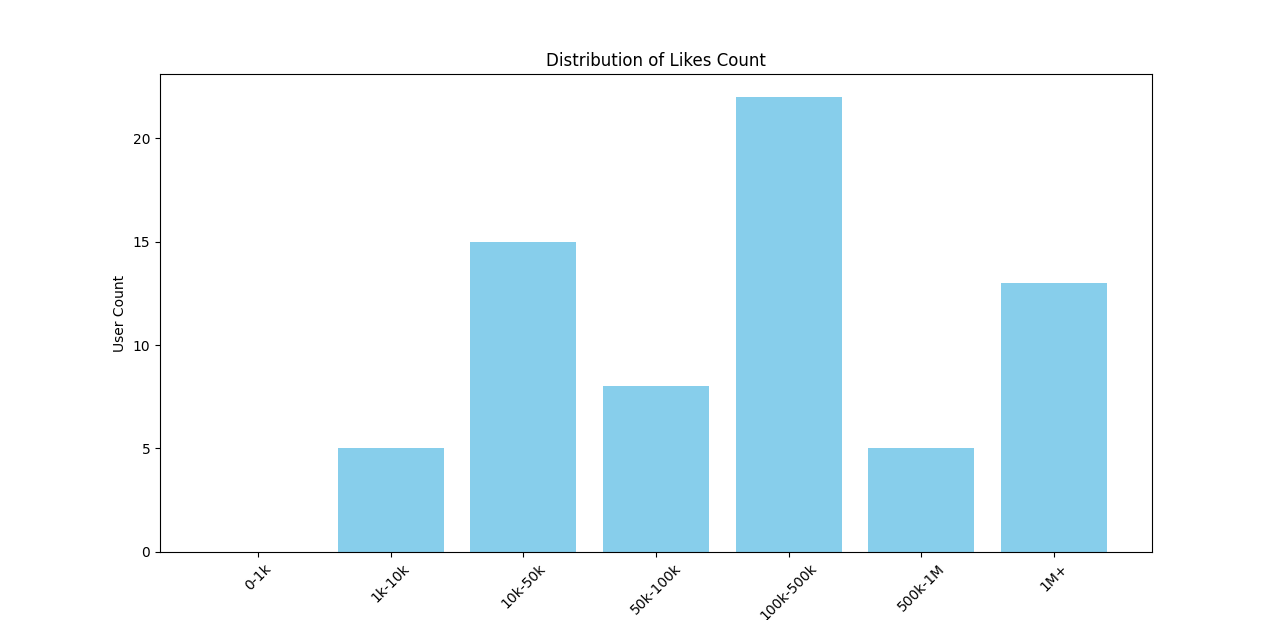}
        \label{fig:likes}
    }
    \hfill
    \subfloat[Videos]{%
        \includegraphics[width=0.32\textwidth]{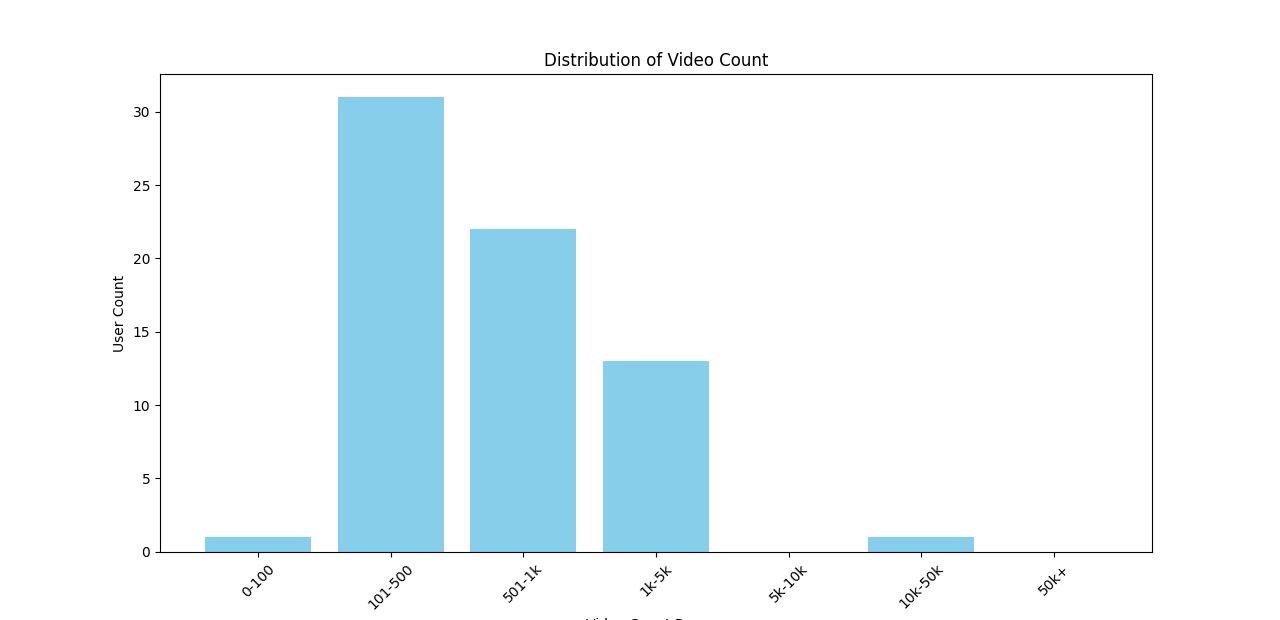}
        \label{fig:videos}
    }
    \hfill
    \subfloat[Followers]{%
        \includegraphics[width=0.32\textwidth]{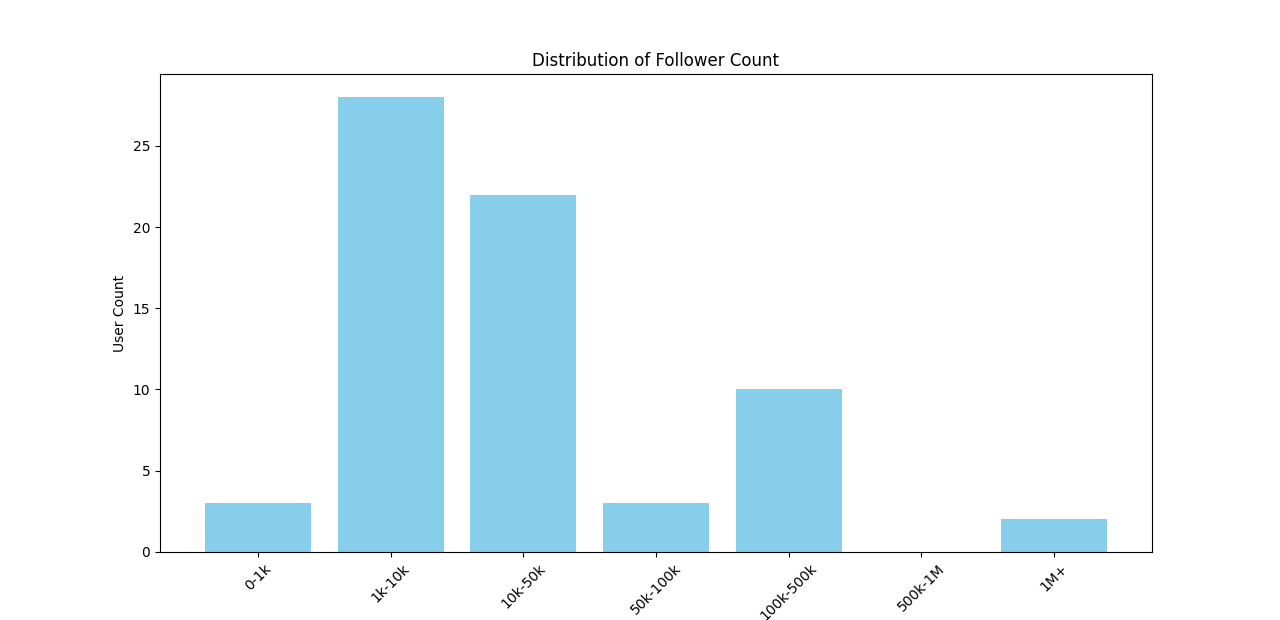}
        \label{fig:followers}
    }
    \caption{Activity metrics of finfluencers: (a) Likes, (b) Videos, and (c) Followers.}
    \label{fig:finfluencers_activity}
\end{figure}

Figures \ref{fig:likes}, \ref{fig:videos}, and \ref{fig:followers} show the numbers of likes, videos and followers, respectively, of those finfluencers. 

%------------------------------------------------------------------
%------------------------------------------------------------------
\subsection{Content, Credibility, and Disclosure in Finfluencer Bios}
\begin{figure}[H]
    \centering
    \includegraphics[width=0.6\textwidth]{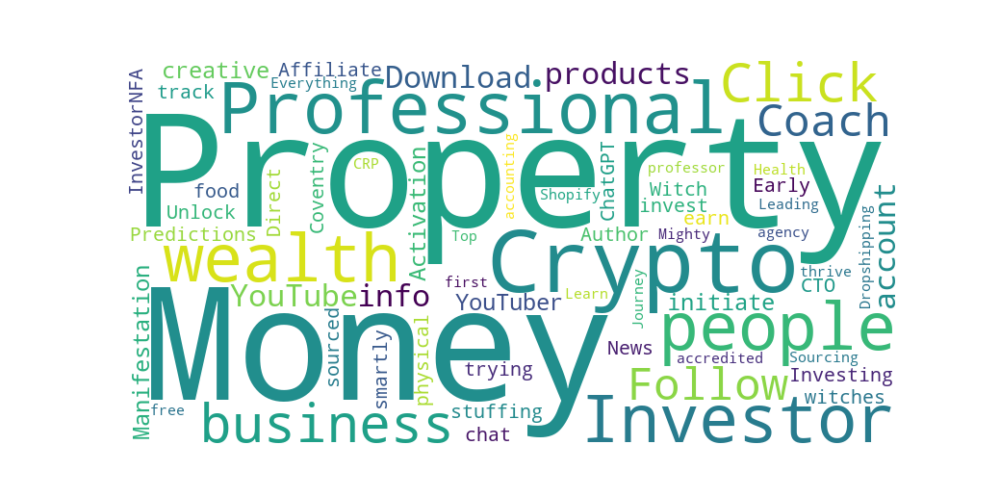}
    \caption{Word cloud showing common terms in finfluencer bios.}
    \label{fig:bio_cloud}
\end{figure}

Figure~\ref{fig:bio_cloud} presents a word cloud illustrating the most common terms used in finfluencers' bios. To preserve user anonymity, any identifying information from the \texttt{username} and \texttt{display\_name} fields was removed prior to the analysis.
Beyond thematic content, we examined credibility signals and disclosure practices. Only two finfluencers included a bio disclaimer stating their content \textit{does not constitute financial advice}, with follower counts of 1K--10K and 10K--50K. Three users were verified, potentially signaling perceived credibility.

The sparse use of disclaimers reveals a gap in disclosure practices. While some may include disclaimers in videos, their absence from bios suggests limited emphasis on transparency at the profile level—despite its importance in building trust in financial communication.

%=============================================
\subsection{Social Graph of Finfluencers}
\begin{figure}[htb]
    \centering
    \includegraphics[width=1\textwidth]{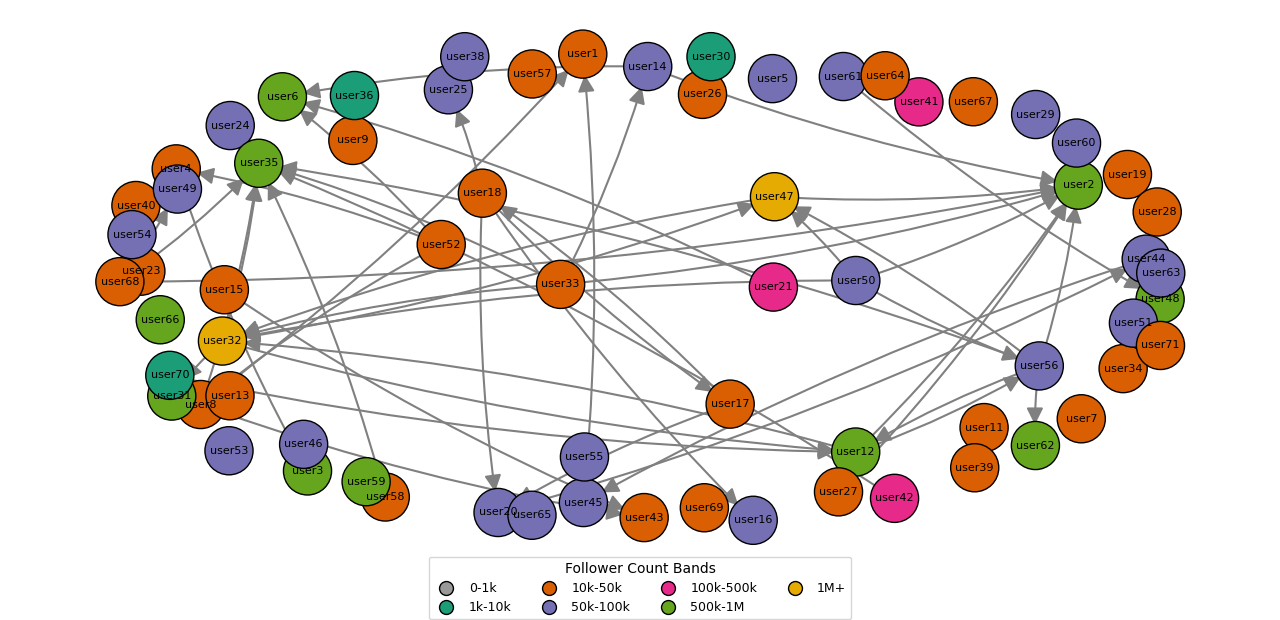}
    \caption{Directed follower network among finfluencers. }
    \label{fig:follower_graph}
\end{figure}

Figure \ref{fig:follower_graph} illustrates how these finfluencers follow one another. To maintain anonymity, we use pseudonyms in the form of \texttt{useri} and follower count ranges. The arrows indicate the direction of the relationship between users.

We compute closeness and betweenness centrality to identify central finfluencers. \texttt{user2} and \texttt{user35} have the highest closeness, suggesting they can reach others efficiently. \texttt{user32} and \texttt{user12} rank highest in betweenness, acting as bridges between groups.

The graph also shows a well-connected structure with several users acting as central hubs, receiving follows from many others. Influencers in the 50k--100k and 10k--50k follower ranges are the most central and followed. Even users with over 100k followers maintain reciprocal connections, indicating mutual engagement across influencer tiers. This suggests a collaborative ecosystem, where mid-tier influencers play a key role in bridging micro- and macro-level influencers.
%Top 5 users by Closeness Centrality:
%user2: 0.1244
%user35: 0.1194
%user31: 0.0672
%user32: 0.0672
%user12: 0.0597

%Top 5 users by Betweenness Centrality:
%user32: 0.0038
%user12: 0.0033
%user56: 0.0020
%user20: 0.0014
%user44: 0.0014

%=============================================

%-------------------------------------------------
\section{Video Data Analysis}
\label{sec:Videos}
In this section, we analyse data related to videos.
\subsection{Duration vs. Engagement }
\begin{figure}[htb]
    \centering
    \includegraphics[width=0.8\textwidth]{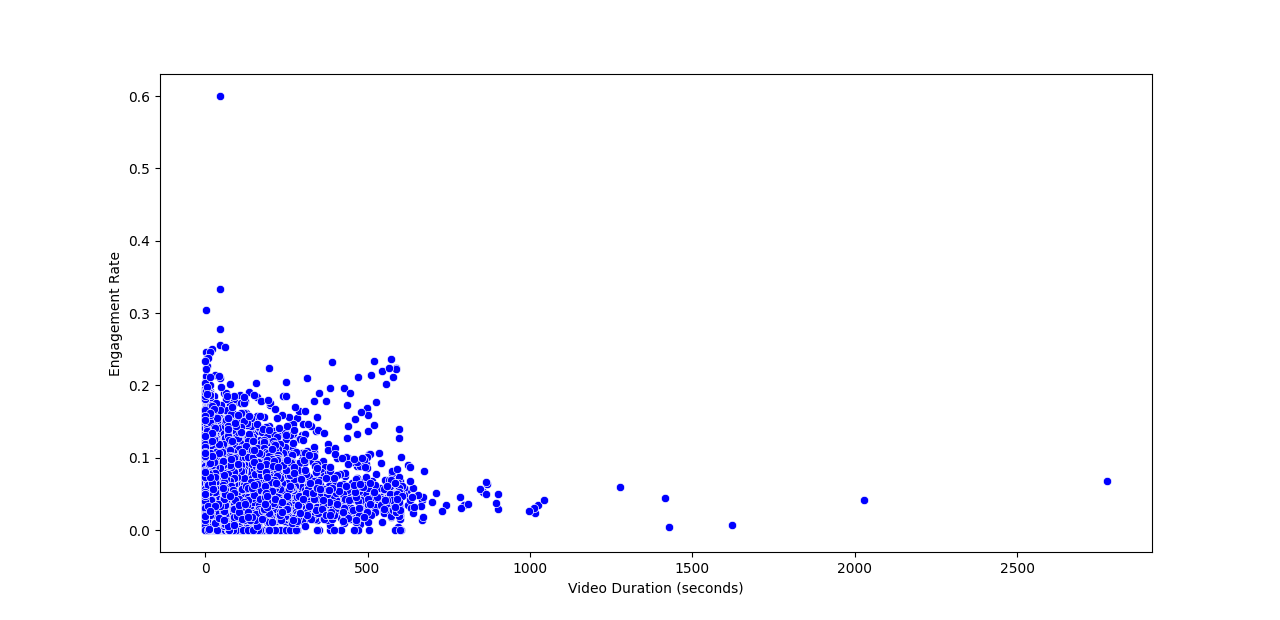}
    \caption{Correlation between video engagement and video duration}
    \label{fig:video_engagement}
\end{figure}

We show in Figure \ref{fig:video_engagement} the correlation between video engagement and video duration. Engagement is calculated as $\frac{\texttt{like\_count} + \texttt{share\_count} + \texttt{comment\_count}}{\texttt{view\_count}}$.

%--------------------------------------------------------------------------------------------------
\subsection{Analysis of Video Hash Tags}
\label{sec:Videohashtags}
Table \ref{tab:top5hashtags} shows the five most frequently used hashtags per month, based on their appearance in TikTok videos. The hashtag \texttt{money} consistently ranked first, highlighting its central role in influencer discussions. Other recurring hashtags like \texttt{crypto}, \texttt{investing}, and \texttt{savemoney} remained in the top five, while \texttt{bargainprices} and \texttt{gooddeals} gained traction in later months, reflecting a growing interest in consumer savings and value.

{%\small
\begin{table}[h!]
\begin{adjustwidth}{-1in}{-1in} 
\centering
\resizebox{1.3\textwidth}{!}{
\begin{tabular}{|c|c|c|c|c|c|c|}
\hline
\textbf{Rank} & \textbf{April} & \textbf{May} & \textbf{June} & \textbf{July} & \textbf{August} & \textbf{September} \\ \hline
1 & money (754) & money (782) & money (616) & money (715) & money (863) & money (721) \\ \hline
2 & forex (382) & crypto (454) & crypto (370) & crypto (400) & savemoney (406) & investing (419) \\ \hline
3 & trading (352) & forex (427) & forex (309) & investing (377) & investing (383) & savemoney (374) \\ \hline
4 & savemoney (325) & trading (385) & savemoney (306) & savemoney (375) & trading (374) & bargainprices (349) \\ \hline
5 & crypto (306) & investing (358) & investing (297) & trading (371) & bargainprices (372) & gooddeals (348) \\ \hline
\end{tabular}
}
\caption{Top 5 Hashtags per Month (April–September 2024).}
\label{tab:top5hashtags}
\end{adjustwidth}
\end{table}
}
%#######################################################################
\begin{comment}
\begin{table}[h!]
    \centering
    % April Table
    \begin{minipage}{0.33\textwidth}
        \centering
        \begin{tabular}{|c|c|}
            \hline
            \multicolumn{2}{|c|}{\bf April} \\ \hline
            \textbf{Hashtag} & \textbf{Count} \\ \hline
            money  &  754  \\
             forex &   382  \\
           trading &   352  \\
         savemoney &  325  \\
            crypto &  306  \\
         investing &  289  \\
         gooddeals &  282   \\
     bargainprices &  282  \\
     cheaperprices &   270  \\
  financialfreedom &  257  \\
  \hline
        \end{tabular}
    \end{minipage}
\end{table}    

\begin{table}[h!]
  % May Table
    \begin{minipage}{0.33\textwidth}
        \centering
        \begin{tabular}{|c|c|}
            \hline
            \multicolumn{2}{|c|}{\bf May} \\ \hline
            \textbf{Hashtag} & \textbf{Count} \\ \hline
money  &  782 \\
crypto &   454 \\
forex   & 427 \\
trading &   385 \\
 investing &   358 \\
savemoney  &  347 \\
passiveincome &   321 \\
bargainprices  &  318 \\
gooddeals   & 316 \\
cheaperprices &   302 \\
\hline
        \end{tabular}
    \end{minipage}
\end{table} 
\end{comment}
%###################################################################################
%##################################################################
%##################################################################
\subsection{Topic Modeling and Thematic Consolidation of Videos}
\label{sec:videothemes}
To uncover latent thematic structures, we applied Latent Dirichlet Allocation (LDA)\cite{LDA2003} with optimized hyperparameters to our dataset of 13,215 videos. The model was trained on a combination of the \texttt{video\_description}, \texttt{hashtag\_names}, and \texttt{voice\_to\_text} fields. Preprocessing included stopword removal, lemmatization, and filtering of general terms.

The LDA model initially identified eight distinct topics, each characterised by a ranked list of keywords. Topic quality was assessed based on coherence scores and manual interpretability. While the coherence score was moderate (0.478, C\textsubscript{v}), qualitative analysis confirmed that the topics were thematically coherent. The number of topics was chosen to balance both coherence and interpretability.
It is noteworthy that Topic 6 exhibited weak influences, with a diverse set of low-weight keywords, indicating that this topic may not represent a coherent or dominant theme within the dataset.

The eight initial topics (along with the top five keywords for each) were:
\begin{sloppypar}
\begin{itemize}
    \item \textit{Topic 0 -- Digital Entrepreneurship \& Online Income:} 
    {entrepreneur}, {digital\_products}, {howtomakemoneyonline},
    {passive\_income}, {digitalmarketingtips}
   
    \item \textit{Topic 1 -- Property Investment \& Wealth Building:} 
    {investment}, property\_investing, {propertyportfolio}, 
    {moneytips}, {wealthbuilding}
    
    \item \textit{Topic 2 -- Forex \& Crypto Trading Education:} 
    {forex\_trading}, {trading}, {tradingforex}, 
    {investment}, {crypto\_trading}
    
    \item \textit{Topic 3 -- Property Buying \& Real Estate Tips:} 
    property\_investing, {buytolet}, {residential}, 
    {deposit}, {isa}
    
    \item \textit{Topic 4 -- Success Mindset \& Startup Motivation:} 
    {mindset}, {liquidity}, {successtips}, 
    {startupstories}, {entrepreneurialjourney}
    
    \item \textit{Topic 5 -- Budgeting \& Personal Finance:} 
    {saving}, {cashstuffingenvelopes}, {debtfree}, 
    {loudbudgeting}, {pennysavingchallenge}
    
    \item \textit{Topic 6 -- Side Hustles \& Work-from-Home:} 
    {sidehustleforbeginners}, {wfh}, {reseller}, 
    {sideincome}, {savemoney}
    
    \item \textit{Topic 7 – Crypto \& Beginner Investing:} 
    {btc}, {crypto}, {crypto\_trading}, 
    {xrp}, {education}
\end{itemize}
\end{sloppypar}

To enhance clarity and reduce redundancy, these were consolidated into four broader themes:
\begin{itemize}
    \item \textit{Topic 0 -- Entrepreneurship \& Side Hustles}: Topics 0, 4, 6
    \item \textit{Topic 1 -- Property Investing}: Topics 1, 3
    \item \textit{Topic 2 -- Active Trading (Forex \& Crypto)}: Topics 2, 7
    \item \textit{Topic 3 -- Saving \& Budgeting}: Topic 5
\end{itemize}

To further evaluate the performance of our topic analysis, we also conducted a separate topic analysis using a BERT-based topic analyzer. The topics identified were very similar to the four merged topics we used to categorize the videos.

Each video was then assigned to its most relevant LDA topic based on keyword overlap in the tokenized text fields. These assignments were then mapped to the merged thematic categories for final analysis.

The topic categorization of the 13,215 videos can be found in Table \ref{tab:topicstats}.
\begin{table}[htb]
\centering
\begin{tabular}{|l|c|}
\hline
\multicolumn{1}{|c|}{\textbf{Topic}} & \textbf{Number of Videos} \\
\hline
Topic 0 -- Entrepreneurship \& Side Hustles & 6447 \\
Topic 1 -- Property Investing & 1670 \\
Topic 2  -- Active Trading (Forex \& Crypto) & 3458 \\
Topic 3   -- Saving \& Budgeting& 1640 \\
\hline
\end{tabular}
\caption{Categorization of the 13,215 videos based on LDA topic assignment.}
\label{tab:topicstats}
\end{table}
%-------------------------------------------------
\subsection{Analysis of Disclaimers in Videos}
\label{sec:disclaim}
To differentiate between finfluencers who clearly encourage their followers to conduct their own research, take independent action, or include other appropriate disclaimers, we analysed the videos for the presence of such disclaimers.

We analysed the fields \texttt{video\_description}, \texttt{voice\_to\_text}, and \texttt{hashtag\_names} for keywords or hashtags associated with disclaimers. A video was classified as containing a disclaimer if any of these fields contained either a keyword or a hashtag associated with disclaimers.
We emphasize that the disclaimer analysis was based on the aforementioned fields, and we did not examine the videos for visual or voice disclaimers. Therefore, some videos lacking \texttt{voice\_to\_text} may still contain disclaimers.

The disclaimer keywords used in the analysis are: \keyw{dyor}, \keyw{not financial advice}, \keyw{educational purposes}, \keyw{consult a financial advisor}, \keyw{do your own research}, \keyw{personal opinion}, \keyw{capital at risk}, \keyw{trading involves risk}, \keyw{results not guaranteed}, \keyw{no liability accepted}, \keyw{crypto is volatile}, \keyw{high risk}, \keyw{not responsible for losses}, \keyw{only invest what you can afford to lose}, \keyw{tread carefully}, \keyw{do not invest more than you can afford to lose}, \keyw{own risk}, \keyw{not responsible}, \keyw{invest what you can lose}, \keyw{educational only}, \keyw{not investment advice},\keyw{seek advice}, \keyw{value can go down}, \keyw{for entertainment only}, \keyw{speculative}, and \keyw{high volatility}.

The hashtags used in the analysis are: \#nfa, \#dyor, \#notfinancialadvice, \#notinvestmentadvice, \#investatyourownrisk, \#entertainmentpurposesonly, \#educationalpurposesonly, \#personalopinion, \#disclaimer, \#tradingatrisk, \#ownresearch, \#highriskinvestment, \#notliable, \#financialresponsibility, \#cryptonfa, \#cryptoinvestmentrisks, \#cryptotipsnotadvice, \#cryptotradingrisks, \#tradingnfa, \#stocktipsnotadvice, \#notarecommendation, \#notadvice, \#investwisely, 
    \#onlyyoucanrisk, \#pleaseconsult, \#entertainmentcontent, \#investsafely, and \#tradingdisclaimer.
    
Table \ref{tab:videodisclaimer} summarizes monthly disclaimer statistics, including the number of distinct finfluencers (\texttt{\# Distinct Finfluencers}) who included disclaimers. Table \ref{tab:hashtag_occurrences} shows the frequency of hashtags in videos containing disclaimer-related tags or keywords.
\begin{table}[htb]
    \centering
    \begin{tabular}{|c|c|c|c|}
    \hline
    \textbf{Month} & \textbf{Total \# Videos} & \textbf{\# Videos with Disclaimers} & \textbf{\# Distinct Finfluencers} \\ \hline
    April & 2203 & 20 & 10 \\ \hline
    May & 2475 & 36 & 14 \\ \hline
    June & 2023 & 34 & 12 \\ \hline
    July & 2205 & 47 & 13 \\ \hline
    August & 2301 & 26 & 11 \\ \hline
    September & 2008 & 45 & 11 \\ \hline
    \end{tabular}
    \caption{Statistics of disclaimers in videos by month.}
    \label{tab:videodisclaimer}
\end{table}

Disclaimer use fluctuated, peaking in July (47 videos) and dipping in April (20). More finfluencers included disclaimers in May (14) and July (13).
\begin{table}[htb]
\centering
\begin{tabular}{|c|l|}
\hline
\textbf{Month} & \multicolumn{1}{|c|}{ \textbf{Top 5 Hashtags \& Their Occurrence}} \\ \hline
April & investingtips (7), crypto (6), money (4), bitcoin (4), investingexplained (4) \\
May & crypto (20), xrp (10), bitcoin (9), investing (7), ripple (7) \\
June & crypto (15), cryptocurrency (8), investing101 (7), xrp (7), bitcoin (6) \\
July & crypto (16), cryptocurrency (13), investing (12), personalfinance (12), bitcoin (10) \\
August & investing (16), cryptocurrency (13), bitcoin (10), crypto (8), money (5) \\
September & cryptocurrency (21), trading (21), crypto (20), bitcoin (20), investing (18) \\
\hline
\end{tabular}
\caption{Top 5 hashtags and their occurrences in videos with disclaimers.}
\label{tab:hashtag_occurrences}
\end{table}
Over six months, the most frequent disclaimer-related hashtags were \#crypto (85), \#investing (60, incl.\ variants like \#investingtips, \#investing101), \#bitcoin (59), \#cryptocurrency (55), and \#trading (21). The dominance of hashtags such as \#crypto, \#investing, and \#bitcoin reflects a strong focus on crypto and investment topics, with variants indicating an educational intent.

The prevalence of disclaimers—especially around volatile assets—suggests rising legal concern and a move toward greater transparency among UK finfluencers.
\subsubsection{Disclaimers by Topic Category.}
%\label{sec:disclaimbycat}
We analyzed the presence of disclaimers across topic categories. Table \ref{tab:disclaimbytopic} shows the number of videos containing some form of disclaimer out of the 215 videos associated with a disclaimer, distributed across four topic categories.

It is notable that Topic 2 — \textit{Active Trading (Forex \& Crypto)} — had the highest number of disclaimers, likely reflecting the influence of recent guidance from regulatory bodies such as the Financial Conduct Authority (FCA).

\begin{table}[htb]
\centering
\begin{tabular}{|l|c|}
\hline
\multicolumn{1}{|c|}{\textbf{Topic}} & \textbf{Number of Videos with Disclaimers} \\
\hline
Topic 0 -- Entrepreneurship \& Side Hustles & 24 \\
Topic 1 -- Property Investing & 21 \\
Topic 2 -- Active Trading (Forex \& Crypto) & 162 \\
Topic 3 -- Saving \& Budgeting & 8 \\
\hline
\end{tabular}
\caption{Distribution of disclaimers by topic category.}
\label{tab:disclaimbytopic}
\end{table}
%It should also be noted that disclaimers were only identified through textual descriptions, hashtags, or the \texttt{voice\_to\_text} fields in videos where that information was available. This does not exclude the possibility that some videos (especially those lacking \texttt{voice\_to\_text} data) may have included disclaimers embedded within the video content itself.

%##################################################################
%###################################################################
\subsubsection{Analysis of Disclaimers by User.}
\begin{figure}[htb]
    \centering
    \subfloat[Likes]{%
        \includegraphics[width=0.32\textwidth, height=4.2cm]{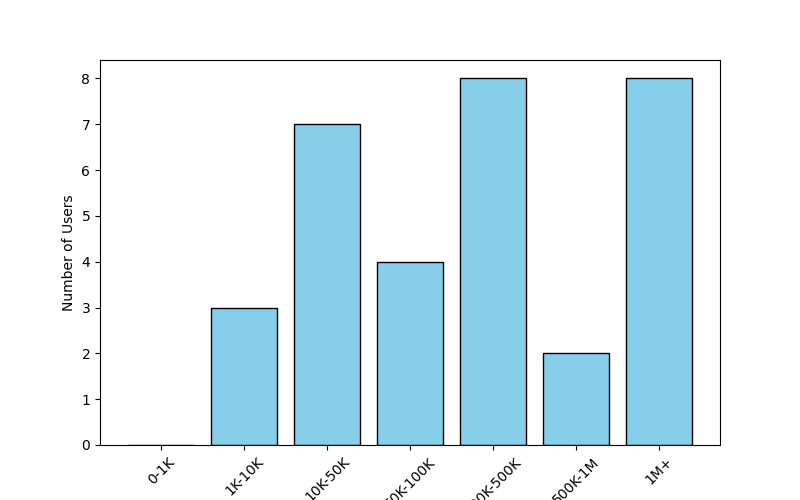}
        \label{fig:disclaimlikes}
    }
    \hfill
    \subfloat[Videos]{%
        \includegraphics[width=0.32\textwidth, height=4.2cm]{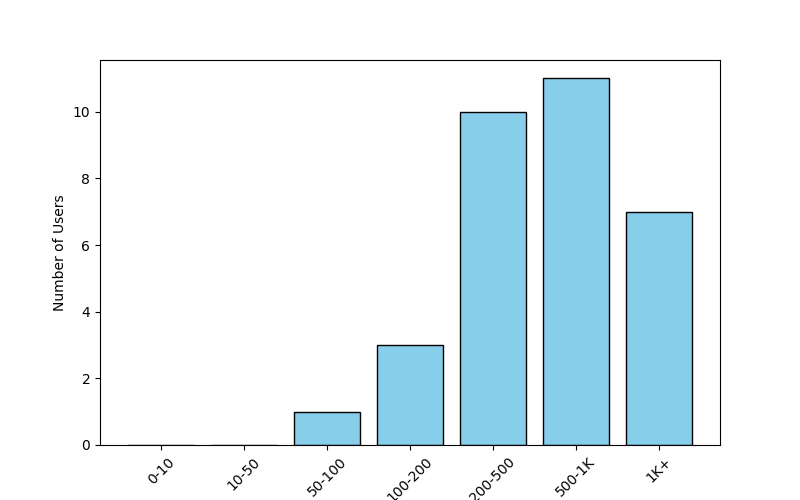}
        \label{fig:disclaimvideos}
    }
    \hfill
    \subfloat[Followers]{%
        \includegraphics[width=0.32\textwidth, height=4.2cm]{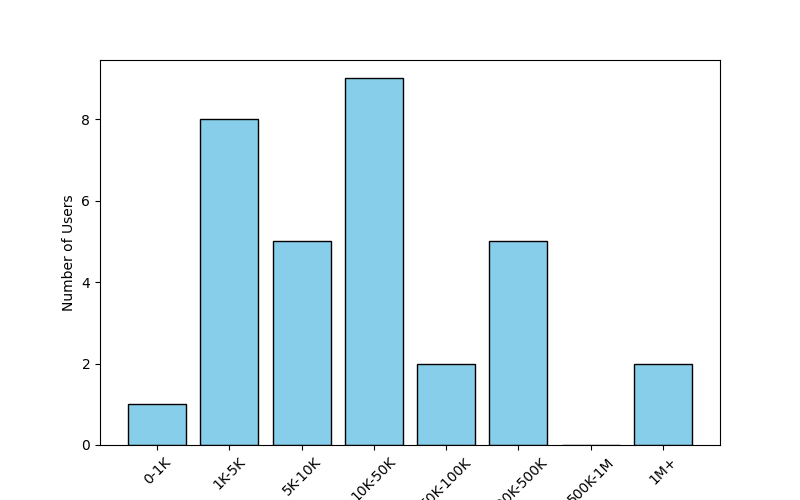}
        \label{fig:disclaimfollowers}
    }
    \caption{With-disclaimer finfluencer metrics: (a) Likes, (b) Videos, (c) Followers.}
    \label{fig:disclaimfinfluencers_activity}
\end{figure}

Figure~\ref{fig:disclaimfinfluencers_activity} presents the number of likes, videos, and followers for finfluencers who included a disclaimer in at least one video. A total of 33 distinct users met this criterion; among them, two also featured a disclaimer in their bio, three had verified accounts, and one account was no longer active.

%Disclaimer detection was based solely on the \texttt{video\_description}, \texttt{hashtag\_names}, and \texttt{voice\_to\_text} fields. The actual video content was not reviewed, so disclaimers presented visually or audibly—especially in videos lacking \texttt{voice\_to\_text}—may not have been captured.
%-------------------------------------------------
\begin{comment}
  
\subsection{Correlation Between Disclaimers and Sentiment Results} \label{sec:disclaimvssentiment}

We analysed the correlation between the presence of disclaimers in videos and the results of sentiment analysis on the corresponding video comments. The table below summarises the number of videos each month whose sentiment analysis was deemed negative, yet included some form of disclaimer. As observed, only a limited number of videos contained a disclaimer and were classified as having negative sentiment. Notably, all these videos had a small number of comments, as shown in the table.

\begin{table}[htb]
    \centering
\begin{tabular}{|c|c|}
\hline
\textbf{Month} & \textbf{\# Videos} \\
\hline
April &  1 ~\\
May & 4 ~\\
June & 1 ~\\
July & 1~\\
August & 2~\\
September & 1~\\
\hline
\end{tabular}
    \caption{Correlation Between Disclaimers and Sentiment Results}
    \label{tab:disclaimvssentiment}
\end{table}

Upon further analysis, we observed that 6 out of the 10 videos with disclaimers and negative sentiment contained hashtags associated with cryptocurrencies. Additionally, all 10 videos had between 1 and 7 comments. This suggests a potential link between crypto-related content, the presence of disclaimers, and negative sentiment in viewer comments. However, the limited number of comments per video indicates that these findings should be interpreted with caution.
\end{comment}
%-------------------------------------------------

\subsection{Sentiment Analysis of Comments}
\label{sec:sencom}
\begin{comment}
We have performed a sentiment analysis of the finfluencers' comments.
%The methodology we used for comment sentiment analysis can be found in Figure~\ref{fig:sencomfig}. 
For text sentiment analysis, we used the Vader SentimentIntensityAnalyzer \cite{Hutto2014} to calculate sentiment scores for cleaned comments. To clean (pre-process) the comments, we removed URLs, mentions, and hashtags. We also normalized repeated characters and kept only alphabetic characters. We finally removed stop words using NLTK \cite{Loper2002}. For each comment in the video, we calculated a sentiment score for the text of the comment and a separate score for the emojis in the comment. We then aggregated the text scores and separately the emoji scores for all comments in the video. The final combined sentiment score is then calculated from the aggregated text (with a weight of 80\%) and the emoji (with a weight of 20\%) score. We then classified the video as either negative, neutral, or positive using its combined score. We also separately calculated the classification based solely on the text score in order to compare the two approaches. We believe that the first approach is more appropriate in this context, as emojis can also signal the user's sentiment.
\end{comment}

We have performed a sentiment analysis of the finfluencers' video comments. We used the Vader SentimentIntensityAnalyzer \cite{Hutto2014} to calculate sentiment scores for cleaned comments. To clean (pre-process) the comments, we removed URLs, mentions, and hashtags. We also normalized repeated characters and kept only alphabetic characters. We finally removed stop words using NLTK 
\cite{Loper2002}. 

Our sentiment analysis process involved calculating  a sentiment score for the text and a separate score for the emojis for each cleaned video's comment. A final sentiment score for each comment was computed as a weighted sum: $0.8 \times \text{text score} + 0.2 \times \text{emoji score}$. This was repeated for all comments, and the average of the combined scores was used to determine the video-level sentiment.
  The video-level sentiment was classified as:
  \begin{itemize}
    \item Negative if the average score is $\geq -1$ and $< -0.05$,
    \item Neutral if the average score is $\geq -0.05$ and $\leq 0.05$,
    \item Positive if the average score is $> 0.05$ and $\leq 1$.
  \end{itemize}

We also separately calculated the classification based solely on the text score in order to compare the two approaches. We believe that the first approach is more appropriate in this context, as emojis can also signal the user's sentiment.

We present the summary of our analysis per month in Table \ref{tab:commentsentiment}. Note that the difference in the monthly video count here and in earlier sections is due to videos with no comments or videos whose comments are not accessible (e.g., they have been deleted). The figure next to each month indicates the number of videos that month with at least one accessible comments.

\begin{comment}
\begin{figure}[htbp]
\centering
\begin{tikzpicture}[node distance=0.6cm and 2cm, font=\small]

% Nodes
\node[io] (input) {Input Video Comments};
\node[process, below=of input] (process) {Preprocess Comments};
\node[process, below=of process] (sentiment) {Sentiment Analysis:  Text Score \& Emoji Score};
\node[note, right=2cm of sentiment] (repeatNote) {Repeated for each comment};
\node[process, below=of sentiment] (combine) {Combine Scores \\ $\text{score} = 0.8 \times \text{text} + 0.2 \times \text{emoji}$};
\node[process, below=of combine] (emojiText) {
\begin{minipage}{7cm}
\centering
\textbf{Video Sentiment Classification:} \\
Neg: $[-1, -0.05)$, Neu: $[-0.05, 0.05]$, Pos: $(0.05, 1]$
\end{minipage}
};
\node[io, below=of emojiText] (output) {Output Video-Level Sentiment};

% Arrows
\draw[line] (input) -- (process);
\draw[line] (process) -- (sentiment);
\draw[line] (sentiment) -- (combine);
\draw[line] (combine) -- (emojiText);
\draw[line] (emojiText) -- (output);

% Dotted line to the repeat note
\draw[dashed, line] (sentiment.east) -- (repeatNote.west);

\end{tikzpicture}

\caption{Our method for Video Sentiment Analysis}
\label{fig:sencomfig}
\end{figure}
\end{comment}

The results indicate a consistently higher number of neutral and positive videos compared to negative ones, suggesting that the tone of financial content on TikTok remains relatively balanced rather than overtly pessimistic. However, the persistently low number of negative videos (e.g., 102 out of 1,465 in April; 39 out of 1,368 in September) may also reflect content creator bias — creators could be deliberately avoiding negativity to maintain engagement or to comply with platform norms.

Interestingly, the ratio of videos containing disclaimers is low overall, typically ranging between 1–9 videos per month per sentiment category. The data show varying trends in comment engagement for videos with disclaimers compared to those without. In some months, videos with disclaimers tend to receive fewer comments on average, while in others, the engagement is either comparable to or exceeds the overall number of comments average. These inconsistencies highlight the complexity of the relationship between disclaimers and comment activity. Given the relatively small number of videos with disclaimers in each month, caution is warranted when drawing conclusions from these results.

Hashtag usage remains broadly consistent across sentiment types, with terms such as \#money, \#crypto, and \#forex dominating both neutral and negative content. This suggests that disclaimers or sentiment do not significantly influence the thematic focus of these videos.

For videos that had negative sentiment and disclaimers, the term \#crypto (including its variant, \#cryptocurrency) was the most frequent hashtag, appearing 9 times. The hashtag \#money followed with 5 mentions. \#forex, \#education, and \#finance each appeared 3 times.

As for negative words (which yield negative sentiment classification) in the same videos, ``ban'' (including ``banned'') appeared 6 times, suggesting recurring themes of restriction. The word ``no'' followed with 5 mentions. Other notable terms included ``lies'', ``scam'', and ``losses'' (each appearing 3 times), reflecting concerns about misinformation, deceit, and financial harm. These results highlight issues of trust, legitimacy, and risk within the content.

Overall, the findings highlight a potential underuse of disclaimers in financial TikTok content and suggest a tension between regulatory caution and audience engagement. Additional investigation is warranted to better understand the role of disclaimers and their impact on viewer interaction.
\begin{table}[h!]
\begin{adjustwidth}{-1in}{-1in} 
\centering
\resizebox{1.3\textwidth}{!}{
\begin{tabular}{|l|c|c|c|c|c|l|}
\hline
\textbf{Mon.} & \textbf{Sentiment} & \textbf{\#V (C/T)}  &  \textbf{VDC}  &  \textbf{Avg. C.}&    \textbf{VD Avg. C.} & \textbf{Top Hashtags} \\
\hline
 Apr    & Negative& 102/118 & 1 & 3.1  & 1 &  money (35), forex (28),  crypto (28), investing (27), trading (22) \\
     (1465)                  & Neutral   & 313/319 &2 &   7.2 & 1.5 &  money (133), forex (85), trading (73), finance (59), crypto (57) \\
%###########################################################
% Updating May
\hline May       & Negative & 110/129 & 5  & 2.4 & 2.4 & forex (35), money (34), trading (32), crypto (30), forextrading (23) \\
                (1609)      & Neutral   & 343/360  & 3 & 5.3 &  3.3 & money (106), crypto (79), forex (66), investing (58), trading (57)\\
\hline
 Jun      & Negative  & 70/91  & 3 & 2.4 & 4.3 &  money (26), crypto (21), forex (21), trading (19), wifimoney (18) \\
            (1339)           & Neutral   & 294/305 & 4 & 4.3  & 1.3 &  money (92), crypto (65), trading (51), investing (49), forex (46) \\
\hline
 Jul    & Negative  & 101/125 & 3 &  2.2  & 3.3 & trading (35), crypto (34), money (33),  forex (32), forextrading (26) \\
                  (1461)      & Neutral  & 329/342 & 9  & 5 & 3.3 &  money (124), crypto (79), investing (58), entrepreneur (58), trading (57)

 \\
\hline
Aug  & Negative & 77/97  & 2  & 3.2&  3 & money (31), trading (25), forex (23), crypto (22), forextrading (21) \\
            (1599)            & Neutral  & 250/259 & 3 &  9.9& 3 &  money (110), trading (42), investing (41), crypto (39), viral (38) \\
\hline
Sept & Negative & 39/52  & 2 &2.3 & 4.5 & money (20), investing (15), trading (12), forex (9), forextrading (7) \\
             (1368)         & Neutral  & 190/196 & 6 & 10.6 & 11.5 &  money (82), investing (52), trading (46), crypto (43), forex (41) \\
\hline
\end{tabular}
}
\caption{\scriptsize Summary of sentiment analysis of videos. The number of videos (\#V) is shown in the format (C/T), where C is the Combined score (i.e., Text and Emoji) and T is Text Only. Hashtag counts are based on the Combined score. \textbf{VDC} denotes the number of videos containing disclaimers. \textbf{Avg. C.} refers to the average number of comments per video, while \textbf{VD Avg. C.} is the average number of comments on videos that include a disclaimer.}
\label{tab:commentsentiment}
\end{adjustwidth}
\end{table}
%###########################################################
%\vspace{-2mm}
\section{Conclusion}
\label{sec:summary}
This study provides a focused analysis of finfluencer content on TikTok in the UK, highlighting dominant themes, engagement trends, sentiment patterns, and the use of disclaimers. Content is largely driven by entrepreneurship and side hustles, followed by active trading topics such as forex and cryptocurrencies. Topic modeling identified four key themes: \textit{Entrepreneurship}, \textit{Property Investing}, \textit{Trading}, and \textit{Budgeting}—reflecting the core narratives in financial discourse on the platform.

Shorter videos, particularly those under five minutes, consistently achieved higher engagement, suggesting that concise content is more effective. Sentiment across videos skewed neutral to positive, with few instances of negative tone. Disclaimers were rare—appearing in under 2\% of videos monthly—and used by only 33 unique finfluencers. Notably, disclaimers were predominantly associated with videos focused on active trading (e.g., forex and cryptocurrency), likely reflecting recent regulatory guidance emphasizing the importance of risk disclosure in financial content. However, their presence showed little effect on engagement or thematic focus, as financial hashtags such as \#crypto and \#bitcoin remained prevalent regardless of disclosure.

These findings suggest creators may prioritize engagement over transparency, potentially minimizing risk warnings to align with platform dynamics. Hashtag trends reflect sustained interest in wealth and value-seeking topics, reinforcing the appeal of optimistic financial messaging.

This analysis is limited to videos with available voice-to-text transcripts and metadata; videos lacking transcribed speech may contain undetected disclaimers. Additionally, exploring alternative sentiment models could yield deeper insights.

Overyall, our findings highlights the need for greater regulatory oversight and further research into how transparency, platform policies, and audience trust shape the evolving role of finfluencers in digital finance. Based on the low prevalence of disclaimers and the emphasis on high-engagement content, platforms and regulators should consider mandating clearer risk disclosures, especially for trading-related content. Additionally, social media platforms could implement standardized content labeling or financial literacy prompts to help users better assess the reliability of financial advice. These recommendations aim to enhance transparency and reduce the risk of harm while preserving the accessibility and educational potential of finfluencer content.

\bibliographystyle{abbrv}
\bibliography{references}

\end{document}